%% file: ms.tex
\documentclass[conference, a4paper]{IEEEtran}
\usepackage{cite}
\usepackage{amsmath,amssymb,amsfonts}
\usepackage{algorithmic}
\usepackage{graphicx}
\usepackage{textcomp}
\usepackage{xcolor}
\def\BibTeX{{\rm B\kern-.05em{\sc i\kern-.025em b}\kern-.08em
    T\kern-.1667em\lower.7ex\hbox{E}\kern-.125emX}}
    
\usepackage{regensky}
\usepackage[firstpage=true]{background}
\usepackage{hyperref}

\SetBgContents{\parbox{\textwidth}{\footnotesize © 2020 IEEE. Personal use of this material is permitted. Permission from IEEE must be
obtained for all other uses, in any current or future media, including
reprinting/republishing this material for advertising or promotional purposes, creating new
collective works, for resale or redistribution to servers or lists, or reuse of any copyrighted
component of this work in other works. DOI: \href{https://doi.org/10.1109/MMSP48831.2020.9287071}{10.1109/MMSP48831.2020.9287071.}}}
\SetBgScale{1}
\SetBgAngle{0}
\SetBgPosition{current page.south}
\SetBgVshift{2.8cm}
\SetBgColor{black}
\SetBgOpacity{1}

\begin{document}

\title{Real-Time Frequency Selective Reconstruction through Register-Based Argmax Calculation
}

\author{\IEEEauthorblockN{Andy Regensky, Simon Grosche, Jürgen Seiler, and André Kaup}
\IEEEauthorblockA{\textit{Multimedia Communications and Signal Processing}\\
\textit{Friedrich-Alexander University Erlangen-Nürnberg (FAU)}\\
Cauerstr. 7, 91058 Erlangen, Germany\\
\{andy.regensky, simon.grosche, juergen.seiler, andre.kaup\}@fau.de}
}

\maketitle

\input{abstract}


\begin{acronym}
	\acro{FSR}{Frequency Selective Reconstruction}
	\acro{FSE}{Frequency Selective Extrapolation}
	\acro{SISD}{Single Instruction Single Data}
	\acro{SIMD}{Single Instruction Multiple Data}
	\acro{MIMD}{Multiple Instruction Multiple Data}
	\acro{SM}{Streaming Multiprocessor}
	\acro{FFT}{Fast Fourier Transform}
	\acro{PSNR}{Peak Signal to Noise Ratio}
	\acro{fps}{frames per second}
	\acro{DFT}{Discrete Fourier Transform}
\end{acronym}

\input{sections/introduction}
\input{sections/fsr}
\input{sections/parallelization}
\input{sections/simulations}
\input{sections/conclusion}

\vspace*{10pt}

\bibliographystyle{IEEEtran}
\bibliography{ms}

\end{document}

%% file: abstract.tex
\begin{abstract}
Frequency Selective Reconstruction (FSR) is a state-of-the-art algorithm for solving diverse image reconstruction tasks, where a subset of pixel values in the image is missing. However, it entails a high computational complexity due to its iterative, blockwise procedure to reconstruct the missing pixel values. Although the complexity of FSR can be considerably decreased by performing its computations in the frequency domain, the reconstruction procedure still takes multiple seconds up to multiple minutes depending on the parameterization. However, FSR has the potential for a massive parallelization greatly improving its reconstruction time. In this paper, we introduce a novel highly parallelized formulation of FSR adapted to the capabilities of modern GPUs and propose a considerably accelerated calculation of the inherent argmax calculation. Altogether, we achieve a 100-fold speed-up, which enables the usage of FSR for real-time applications.
\end{abstract}

%% file: sections/introduction.tex
\vspace*{6pt}

\section{Introduction}

Efficient image reconstruction algorithms are in high demand due to their broad applicability to diverse signal processing tasks, such as error concealment \cite{Zhai2010}, image inpain-\mbox{ting \cite{Guillemot2014}} or resolution enhancement \cite{Seiler2015}. \ac{FSR} \cite{Seiler2015} has been introduced as an image reconstruction algorithm in the context of image resolution enhancement and employs non-regular sampling to increase the quality of the reconstructed high resolution image. By tailoring the image acquisition process and the reconstruction algorithm to each other, accurate reconstructions can be achieved. In so-called quarter sampling \cite{Schoberl2011}, a higher resolution is achieved without increasing the number of pixels on the sensor by non-regularly covering $3/4$ of each pixel and reconstructing the high resolution image using \ac{FSR}. This yields a better quality of the high resolution image than the extrapolation of conventional image sensor data while reducing the cost compared to increasing the resolution of the image sensor.

The main problem of most modern reconstruction algorithms, be it in the context of image signals or in the related field of Compressed Sensing \cite{Donoho2006}, \cite{Candes2006} in general, is their high computational complexity and the resulting long execution times. For \ac{FSR}, different approaches have been followed to reduce the overall reconstruction time on the CPU. In \cite{Genser2018}, \ac{FSR} has been adapted to perform real-valued operations exclusively by reconstructing the image signal in the Hartley domain. In \cite{Genser2018a}, constraints have been imposed on \ac{FSR} in order to reduce the total number of operations that is required to reconstruct an image, yet, potentially impairing the overall reconstruction quality. In \cite{Seiler2011}, extensive pre-computations are performed to speed up the iterative reconstruction procedure while increasing the memory cost of the algorithm.

On the other hand, many algorithms have been adapted to exploit the massive parallelization capabilities of modern GPUs. In \cite{Andrecut2008}, the Matching Pursuit algorithm \cite{Mallat1993} for sparse signal recovery has been sped up by performing the involved matrix-vector operations on the GPU. In \cite{Fang2011}, this idea was extended to Orthogonal Matching Pursuit (OMP) \cite{Pati}. In \cite{Dai2014}, the idea is furthermore extended to the 2D-OMP \cite{Fang2012}, where a custom implementation of the required $\text{argmax}$ operation on the GPU is employed, that exploits the increased speed of shared memory compared to global device memory.

\ac{FSR} allows for a massive parallelization on the GPU as well. In this paper, we introduce a parallelization of the \ac{FSR} algorithm on multiple levels and propose a highly effective, high speed argmax calculation. Other than \cite{Dai2014}, which uses shared memory for thread cooperation, we exploit the rapid register access between neighbouring threads on the GPU. By reducing global and shared memory access to a minimum, a considerable speed-up of the overall reconstruction procedure is achieved.

This paper is organized as follows: Section II shortly recaps the \ac{FSR} algorithm forming the basis of this work. \mbox{Section \ref{sec:gpu-fsr}} describes the proposed parallelization of FSR for GPUs and explains the novel register-based argmax calculation in detail. Section IV compares the execution time of the proposed highly parallelized GPU implementation of FSR to its CPU counterpart, as well as a shared memory based GPU implementation, and evaluates the achievable framerates for different resolutions. Finally, \mbox{Section \ref{sec:conclusion}} concludes this paper.

%% file: sections/fsr.tex
\section{Frequency Selective Reconstruction}

\ac{FSR} is a blockwise reconstruction algorithm, that subdivides the image into neighboring target blocks of \mbox{size $B \times B$} pixels and reconstructs each block independently.
During the iterative reconstruction procedure, it takes a neighborhood of $L$ pixels to all sides of the target block into account to take advantage of additional information. The resulting block of \mbox{size $S \times S = (B + 2L) \times (B + 2L)$} pixels is called the support block. \ac{FSR} builds a model $g[m,n]$ for each support block as a weighted superposition of 2D \mbox{\ac{DFT}} basis images $\varphi_{kl}$ \cite{Seiler2015}, \cite{Kaup2005},
\begin{align}
	g[m,n] = \sum_{k=0}^{S-1} \sum_{l=0}^{S-1} c_{kl}\varphi_{kl}[m,n]
\end{align}
with pixel positions $(m,n)$ and frequency components $(k,l)$.
It is capable of reconstructing images where a large subset of pixel values are missing. The sampled signal $\tilde{f}[m,n]$ for a support block is obtained from the support block \mbox{signal $f[m,n]$} as
\begin{align}
	\tilde{f}[m,n] = \begin{cases}
		f[m,n] & \text{for}\ (m,n) \in \mathcal{K}, \\
		0 & \text{otherwise},
	\end{cases}
\end{align}
where the set $\mathcal{K}$ subsumes the indices of all sampled pixels.
It follows an iterative approach to generate the model $g[m,n]$, where in each iteration $\nu$, the coefficient $c_{kl}$ of exactly one basis \mbox{image $\varphi_{kl}[m,n]$} is changed. It is selected such that the weighted residual energy
\begin{align}
	E_w^{(\nu)} = \sum_{m,n} | r^{(\nu)}[m,n] |^2 w[m,n]
\end{align}
is minimized. Thereby, the residual $r^{(\nu)}[m,n]$ describes the difference between the sampled image $\tilde{f}[m,n]$ and the \mbox{model $g[m,n]$} in \mbox{iteration $\nu$} as
\begin{align}
	r^{(\nu)}[m,n] = \tilde{f}[m,n] - g^{(\nu)}[m,n],
\end{align}
and $w[m,n]$ describes the spatial weighting
\begin{align}
	w[m,n] = \begin{cases}
	\hat{\rho}^{\sqrt{\left(m-\frac{S-1}{2}\right)^2 + \left(n-\frac{S-1}{2}\right)^2}} & \text{for}\ (m,n) \in \mathcal{K}, \\
	0 & \text{otherwise},
	\end{cases}
\end{align}
where the scalar decay factor $\hat{\rho}$ controls the speed of decay of the exponentially decreasing weight \cite{Seiler2015}, \cite{Kaup2005}. Unlike \cite{Seiler2015}, we refrain from taking into account already reconstructed pixel values in exchange for a more efficient parallelization of the overall algorithm.
The spatial weighting is employed to ignore unknown samples $(m,n){\ \notin\ }\mathcal{K}$ during the calculation of the weighted residual energy and to assign an exponentially decreasing weight to pixels further away from the block center.

As each basis image $\varphi_{kl}[m,n]$ corresponds to exactly one frequency component, the algorithm can be efficiently performed in the frequency domain. By describing the weighted \mbox{residual $r_w^{(\nu)}[m,n]$} as
\begin{align}
	r_w^{(\nu)}[m,n] = r^{(\nu)}[m,n]\cdot w[m,n],
\end{align}
and formulating the model $g^{(\nu)}[m,n]$, the weighted \mbox{residual $r_w^{(\nu)}[m,n]$} and the spatial weighting $w[m,n]$ in the frequency domain, we \mbox{get $G^{(\nu)}[k,l]$, $R_w^{(\nu)}[k,l]$ and $W[k,l]$}, respectively. The initial model $G^{(0)}[k,l]$ is set to $0$.
\mbox{With $\mathcal{F}_2$} denoting the 2D \ac{DFT}, the weighted \mbox{residual $R_w[k,l]$} is therefore initialized to
\begin{align}
	R_w^{(0)}[k,l] = \mathcal{F}_2\{\tilde{f}[m,n]\cdot w[m,n]\}. \label{fsr:eq:residual_init}
\end{align}
In each iteration, the basis image $\varphi_{kl}[m,n]$ is selected that reduces the weighted residual energy $E_w^{(\nu)}$ the most, so that the frequency index $(u,v)^{(\nu)}$ of the selected basis \mbox{image $\varphi_{uv}[m,n]$} in iteration $\nu$ is obtained as
\begin{align}
	(u,v)^{(\nu)} = \text{argmax}_{(k,l)}\ \left(w_f[k,l]\cdot \vert R_w^{(\nu)}[k,l] \vert^2 \right). \label{fsr:eq:selection_argmax}
\end{align}
Thereby, the frequency weighting $w_f[k,l]$ is incorporated to favor lower frequency basis images over higher ones \cite{Seiler2015}. It is defined as
\begin{align}
	w_f[k,l] = \left(1-\sqrt{2}\sqrt{\frac{\tilde{k}^2}{S^2} + \frac{\tilde{l}^2}{S^2}} \right)^2
\end{align}
with $\tilde{k} = \frac{S}{2} - \vert k-\frac{S}{2} \vert$ and $\tilde{l} = \frac{S}{2} - \vert l-\frac{S}{2} \vert$.
The projection coefficient $p_{uv}^{(\nu)}$ is chosen such that the selected basis \mbox{image $\varphi_{uv}[m,n]$} maximally reduces the weighted residual \mbox{energy $E_w^{(\nu)}$} resulting in
\begin{align}
	p_{uv}^{(\nu)} = \frac{R_w^{(\nu)}[u,v]}{W[0,0]}. \label{fsr:eq:projection_coeff}
\end{align}
Once the frequency component $(u,v)^{(\nu)}$ to be updated has been determined, and its projection coefficient $p_{uv}^{(\nu)}$ has been calculated, the model $G[k,l]$ and the weighted \mbox{residual $R_w[k,l]$} are updated accordingly 
\begin{align}
	G^{(\nu+1)}[u,v] &= G^{(\nu)}[u,v] + \gamma p_{uv}^{(\nu)} S^2, \label{fsr:eq:model_update} \\[0.5em]
	R_w^{(\nu+1)}[k,l] &= R_w^{(\nu)}[k,l] - \gamma p_{uv}^{(\nu)} W[k-u, l-v]\ \forall\ (k,l). \label{fsr:eq:residual_update}
\end{align}
Thereby, the model $G[k,l]$ needs to be updated for the selected frequency component $(u,v)^{(\nu)}$ only, whereas the weighted residual $R_w[k,l]$ needs to be updated for all of the $S^2$ frequency components $(k,l)$. The scalar orthogonality deficiency compensation factor $0 < \gamma \leq 1$ is introduced to reduce the interference between the different basis images \cite{Seiler2015}. This procedure of selecting a basis function according to \eqref{fsr:eq:selection_argmax}, computing the projection coefficient according to \eqref{fsr:eq:projection_coeff}, and updating the model and the residual according to \eqref{fsr:eq:model_update} and \eqref{fsr:eq:residual_update}, respectively, is then repeated for a fixed number of iterations.

Eventually, the final model $G[k,l]$ is transformed back into the spatial domain to obtain $g[m,n]$. The final support block reconstruction $\hat{f}[m,n]$ is obtained by taking over the sampled support block signal $\tilde{f}[m,n]$ for known samples, and taking over the final model $g[m,n]$ for unknown samples.
To end the reconstruction of the current block, the reconstructed target block signal is extracted from the reconstructed support block signal $\hat{f}[m,n]$ and placed at the corresponding target block in the global reconstruction.
Please refer to \cite{Seiler2015} for a more detailed explanation of \ac{FSR} and its parameters.

%% file: sections/parallelization.tex
\section{Frequency Selective Reconstruction\\ on the GPU} \label{sec:gpu-fsr}

\ac{FSR} in the frequency domain shows great potential for large speed improvements through a massive parallelization of the internal processing procedure. Especially on GPUs, which are specifically designed for running many lightweight tasks simultaneously, a careful design of the parallelized \ac{FSR} algorithm promises a huge gain in processing speed. In this section, we describe the applied GPU thread model and explain the highly parallelized \ac{FSR} algorithm that is specifically adapted to it. Furthermore, we take a deeper look at the novel register-based argmax calculation on the GPU, which serves as a major accelerator of the overall reconstruction procedure.

\subsection{Parallelization of FSR} \label{subsec:parallel-fsr}

Parallelizing \ac{FSR} requires a certain understanding of the thread model that the applied GPU uses. The thread model describes how data-parallel and task-parallel workloads are subdivided on the device, and hence, understanding its basics is essential to be able to design an efficient GPU algorithm. As Nvidia\textsuperscript\textregistered's CUDA\textsuperscript\textregistered\ \cite{NVIDIACorporation2019}, \cite{Nickolls2008} programming language enjoys high popularity in the scientific community and due to the broad availability of compatible GPU devices, we employ the CUDA thread model in our work. Note though, that competitor GPU manufacturers commonly use similar models with slightly varying names.

\begin{figure}[t]
\centering
\includegraphics[width=0.92\columnwidth]{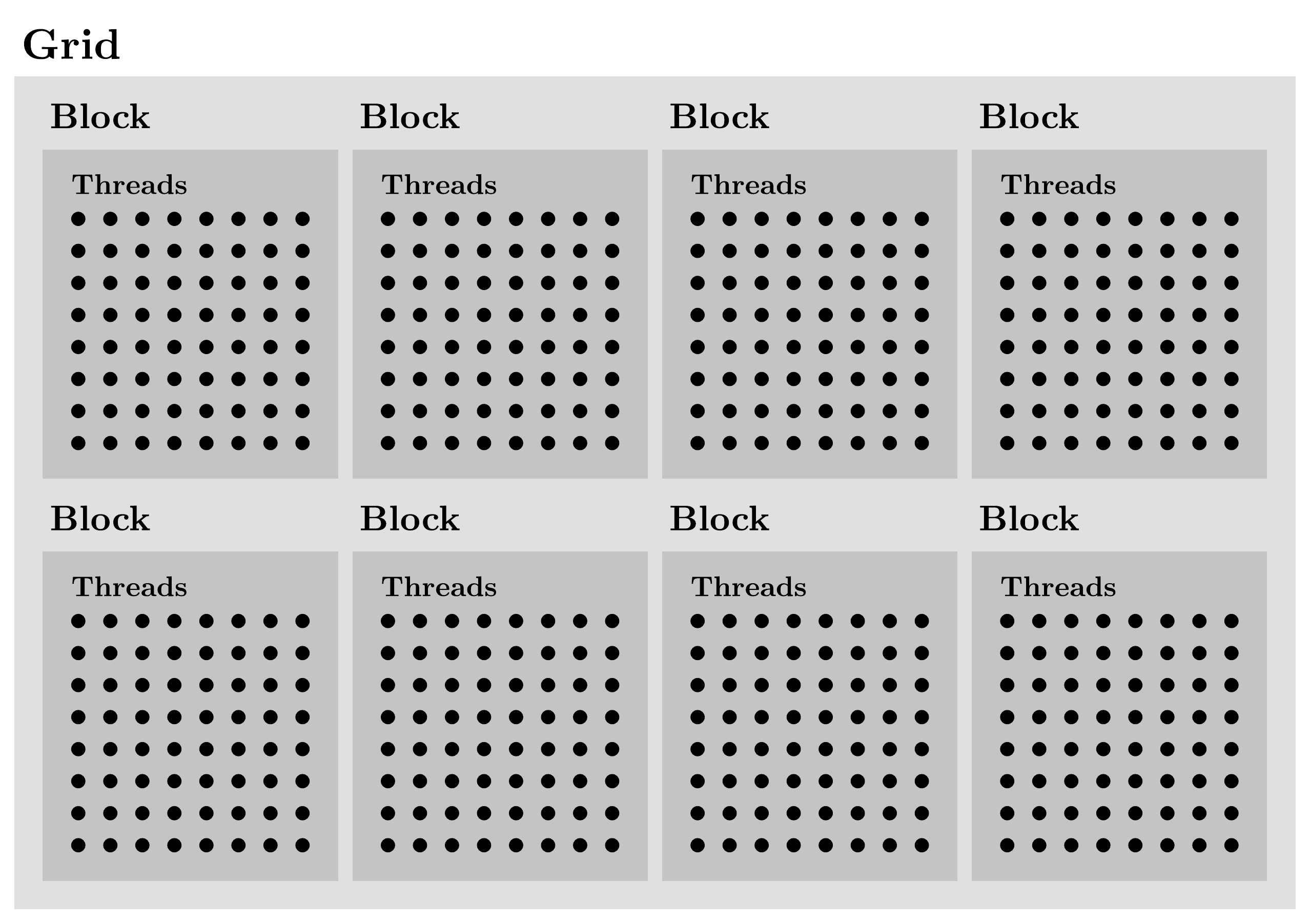}
\caption{Schematic of the employed thread model.}
\label{figure:thread-model}
\end{figure}

Fig. \ref{figure:thread-model} shows the CUDA thread model \cite{NVIDIACorporation2019} where a grid consists of multiple blocks and each block consists of multiple threads. Thereby, all blocks in a grid can run independent from each other, whereas all threads in a block are executing the same program on multiple data elements. All threads in a block run in a shared memory space, i.e., on the same \ac{SM} \cite{NVIDIACorporation2019}, and can interact with each other. Threads in different blocks can not natively interact with each other and data transfer among them requires the involvement of comparably slow global memory. With this basic understanding of the thread model, an efficient way to parallelize \ac{FSR} can be derived.

The parallelization of \ac{FSR} happens in two stages: a task-parallel stage and a data-parallel stage. The task-parallel stage is thereby established similarly to how one would parallelize the reconstruction process on a multi-core CPU. As \ac{FSR} is a blockwise procedure and all blocks can be reconstructed independent from each other, parallelizing the reconstruction of all blocks is an obvious solution. This means that the reconstruction of each support block is assigned to a single thread block on the GPU, each.
For an image of size $Y \times X$, a number of 
\begin{align}
	N = \left\lceil \frac{Y}{B} \right\rceil \cdot \left\lceil \frac{X}{B} \right\rceil
\end{align}
support blocks need to be reconstructed, which results in $N$ thread blocks being employed for the reconstruction of the image. Note, that one can create more thread blocks than there are \ac{SM}s on the GPU, as they are not required to run simultaneously and one \ac{SM} can hold multiple thread blocks.
The data-parallel stage follows a more sophisticated layout for the reconstruction of one support block using \ac{FSR}. In general, the parallelization can be interpreted as each thread being responsible for one pixel of the support block. This means, that each thread is dedicated to a pixel position $(m,n)$ in the spatial domain, or a frequency component $(k,l)$ in the frequency domain, and a support block is reconstructed \mbox{by $S^2$ threads} in parallel. Starting from the regarded sampled support block signal $\tilde{f}[m,n]$, the \mbox{residual \eqref{fsr:eq:residual_init}} is initialized by first multiplying each pixel of the sampled \mbox{signal $\tilde{f}[m,n]$} with the spatial weighting $w[m,n]$ simultaneously, where each thread performs a single multiplication, and then executing a 2D FFT using all threads to obtain the residual in the frequency domain. The FFT algorithm is also executed on the GPU but not investigated further since the optimized and tested \textit{cuFFT} library \cite{NVIDIACorporation2019a} is used for all FFT and IFFT invocations on the GPU. The model $G[k,l]$ is initialized with all threads setting the value of their respective frequency component \mbox{to $0$}. From here, the iterative procedure begins. In each iteration, a basis image is selected, its optimal projection coefficient is calculated, and the model and the residual are updated accordingly. For the selection of the basis image \eqref{fsr:eq:selection_argmax}, each thread calculates the objective value for the frequency component $(k,l)$ it is assigned to. Then, all threads in the regarded thread block are cooperating to find the index of the maximum frequency component \mbox{in $\lceil 2\log_2(S) \rceil$} steps. The thread cooperative argmax calculation is explained in detail in Section \ref{subsec:argmax}. Once the basis image to be added has been found, its projection coefficient \eqref{fsr:eq:projection_coeff}
has to be computed. For this computation, only one thread is involved as the projection coefficient is only required for the selected basis image. The same is true for the update of the model \eqref{fsr:eq:model_update},
so that this computation is performed by a single thread, as well. The update of the residual \eqref{fsr:eq:residual_update}
can be done by all threads simultaneously, where a thread dedicated to frequency component $(k,l)$ accesses the corresponding \mbox{value $W[k-u, l-v] = W[i,j]$} from the Fourier transformed spatial weighting with
\begin{align}
i &= \begin{cases}
	k-u & k-u \geq 0, \\
	S+k-u & \text{otherwise},
\end{cases}\\[1em]
j &= \begin{cases}
	l-v & l-v \geq 0, \\
	S+l-v & \text{otherwise}.
\end{cases}%
\end{align}
After the desired number of iterations has been executed, the final model $G[k,l]$ is transformed back into the spatial domain using the IFFT provided by the cuFFT library. Eventually, each thread writes its assigned pixel value from the reconstructed support block signal $g[m,n]$ to the corresponding pixel position in the final reconstructed image.

\subsection{Register-based argmax calculation} \label{subsec:argmax}

On NVIDIA GPUs, threads are scheduled to run in groups called warps \cite{NVIDIACorporation2019}. Thereby, a warp consists of 32 threads on all current Nvidia GPUs. Within each warp, all threads perform the same instructions on different data elements simultaneously.
The CUDA programming model provides functionalities for threads within the same warp to read out each others registers. Direct register access is extremely fast and allows for a high-performance implementation of the argmax calculation that occurs in the iterations of \ac{FSR}.

As a first step, all threads of the current block compute the objective value for their corresponding frequency \mbox{component $(k,l)$}
\begin{align}
	\text{obj}[k,l] = w_f[k,l]\cdot \vert R_w[k,l] \vert^2,
\end{align}
which is stored in a register \texttt{objective} in each thread. Furthermore, each thread stores its frequency component $(k,l)$ in registers \texttt{index\_k} and \texttt{index\_l}.

The argmax calculation is then performed in two phases. In the first phase, the argmax is searched within each warp of the current block independently. Thereby, the threads within the current block are assigned to warps in a consecutive fashion.  In the second phase, the results from the warp-internal first phase are further processed by a single warp to find the argmax result of the overall thread block. Both, the first phase and the second phase perform the same underlying iterative, thread-cooperative procedure to rapidly find the argmax result by optimally utilizing the available GPU resources. This procedure consists of three steps that are executed simultaneously by all threads in a warp:

\begin{enumerate}
	\item A thread at index \texttt{i} reads the register value \texttt{objective} of the offset thread at index $\texttt{i} + \texttt{offset}$ and stores it in a register \texttt{objective\_offset}. Preceeding the first iteration, the offset is initialized to half the number of threads per warp.
	\item If the objective value of the offset thread is larger than the objective value of the current thread ($\texttt{objective\_offset} > \texttt{objective}$), the current thread replaces its register values \texttt{objective}, \texttt{index\_k} and \texttt{index\_l} with the corresponding register values from the offset thread.
	\item While the offset is larger than $1$ ($\texttt{offset} > 1$), the offset is halved and the procedure continues with Step 1. Otherwise, the warp-internal argmax calculation finishes and the argmax result is guaranteed to reside within the first thread of the warp.
\end{enumerate}

\begin{figure}
\centering
\includegraphics[width=\columnwidth]{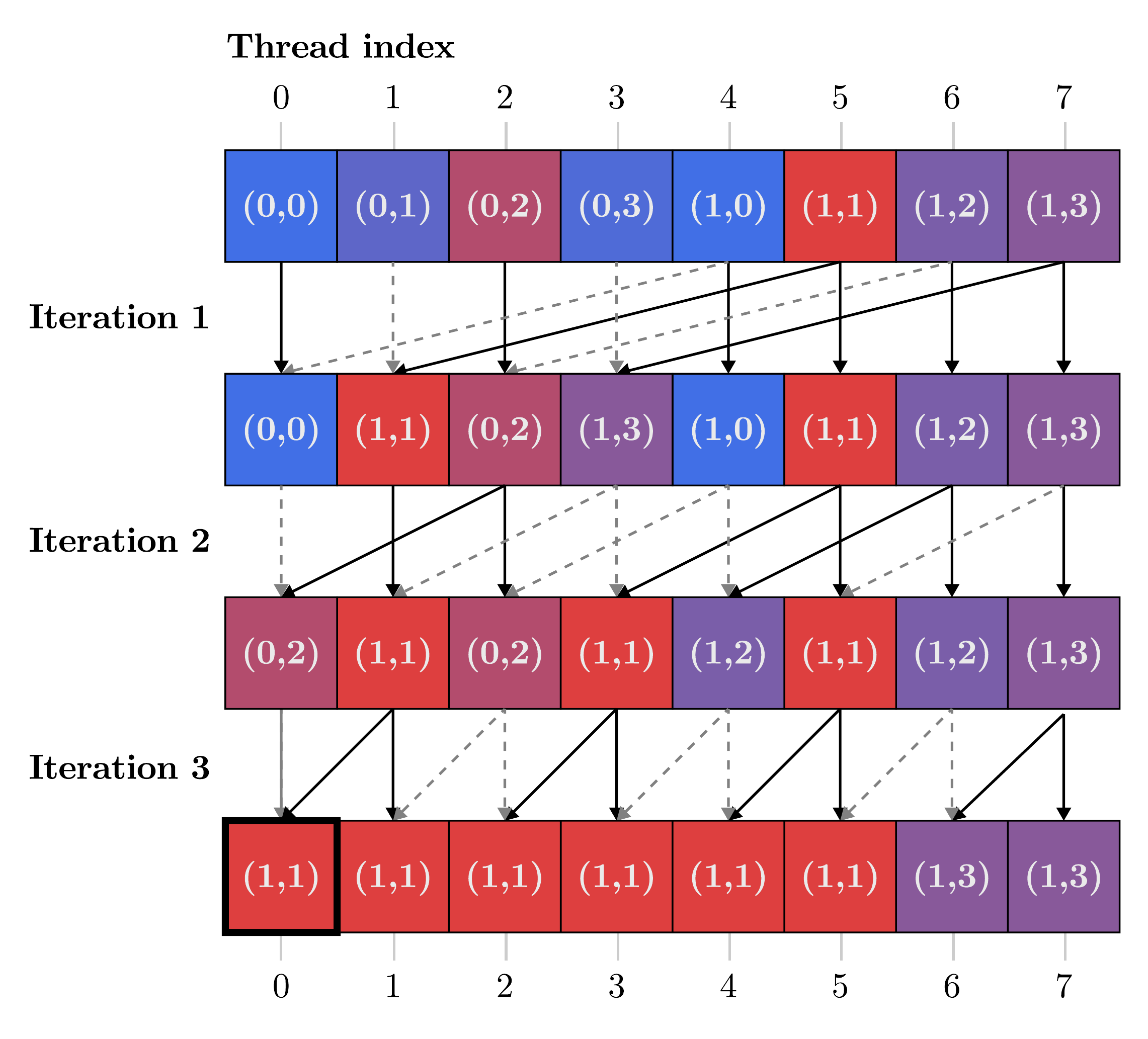}
\caption{Schematic of the argmax calculation within a single warp. Simplified representation with 8 threads per warp. Cell values correspond to the frequency component. Colors represent the objective value, where blue represents low values and red represents high values. In each iteration, each thread compares its own objective value (vertical arrow) with the offset thread's objective value (angled arrow) and takes over the register values from the thread with the higher objective value (solid black arrow). After the last iteration, the result of the argmax calculation lies in the first thread of the warp (black box).}
\label{fig:warp-argmax}
\end{figure}

The described procedure is depicted in Fig. \ref{fig:warp-argmax}, where a simplified argmax calculation with 8 threads per warp is performed. Note, that if a thread requests an out-of-bound thread register, its own register value is returned instead. Thereby, a thread register is considered out-of-bounds if it is not a member of the current warp.

Once the first phase completes, each warp's first thread contains the warp-internal argmax result (maximum objective value and corresponding frequency component) and stores it in shared memory for later access.
In the second phase, the argmax of the overall block is calculated by evaluating the argmax among all warp-internal results from the first phase. As CUDA allows a maximum number of 1024 threads per block, there are at most 32 warps per block. Hence, there exist at most 32 warp-internal argmax results in shared memory after finishing the first phase. Consequently, in the second phase, a single warp can perform the argmax calculation using the described highly efficient thread-cooperative procedure. This is achieved by each thread reading the results of one warp-internal argmax calculation from shared memory and storing the corresponding values in its local registers \texttt{objective}, \texttt{index\_k} and \texttt{index\_l}. The iterative, thread-cooperative procedure is then performed as described above. Finally, independent of the chosen \mbox{blocksize ($S^2 \leq 1024$)}, the final argmax result will always reside in the first thread of the warp that executed the second phase of the described two-phase procedure. From there, the frequency \mbox{component $(k,l)$} corresponding to the maximum objective value can be extracted and the \ac{FSR} algorithm can continue as described in \mbox{Section \ref{subsec:parallel-fsr}}. Due to the minimal access of shared \mbox{memory (global} memory is not involved), the register-based argmax calculation is extremely fast and yields a considerable speed-up of the overall reconstruction procedure.

%% file: sections/simulations.tex
\section{Simulations}

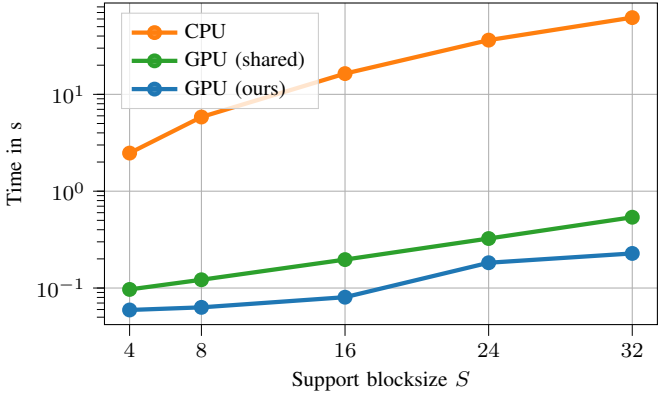
\begin{figure}[t]
	\setlength\figurewidth{\columnwidth}
	\setlength\figureheight{.66\columnwidth}
	\centering
	{\footnotesize\input{tikz/time_gpu_cpu}}
	\caption{Comparison of the execution time of the FSR on the CPU and the GPU with a shared memory and our register-based argmax approach.}
	\label{simulations:fig:time_gpu_cpu}
\end{figure}

To evaluate the performance of the proposed parallelization of \ac{FSR} for GPUs including the novel register-based argmax calculation, the reconstruction procedure on the GPU is compared to the reconstruction procedure on the CPU in terms of quality and speed. To realize a fair comparison, a highly optimized CPU implementation in C++ is employed. All CPU computations are performed on one core of an Intel\textsuperscript{\textregistered} Xeon\textsuperscript{\textregistered} E5 2690 @ 2.6 GHz. Simulations that use the GPU implementation run on an Nvidia GeForce\textsuperscript{\textregistered} RTX 2080 Ti with 11 GB of global device memory featurng 4608 cores. Thereby, code that runs on the GPU is written in Nvidias CUDA programming language and all parts that require CPU interaction (memory transfer to/from the GPU, starting the reconstruction procedure) are written in C++\footnote{The source code for all evaluated implementations of FSR is publicly available at https://gitlab.lms.tf.fau.de/lms/gpu-fsr}. The cuFFT library \cite{NVIDIACorporation2019a} is used for the necessary FFT calculations on the GPU. Execution time and Peak Signal to Noise \mbox{Ratio (PSNR)} are averaged over 100 grayscale images from the TECNICK image \mbox{dataset \cite{Asuni2013}} for a given parameterization. If not stated otherwise, the images are processed at their original resolution of $1200 \times 1200$ pixels. The target blocksize is set \mbox{to $B = 4$} pixels, the spatial decay factor to $\hat{\rho} = 0.7$ and the orthogonality deficiency compensation factor to $\gamma{\ =\ }0.5$. The sampled \mbox{image $\tilde{f}$} is obtained through a quarter sampling sensor \cite{Seiler2015}, which is emulated by sampling a random pixel in each neighboring $2 \times 2$ block of the original image $f$.

\begin{figure}[t]
\centering
\includegraphics[width=\columnwidth]{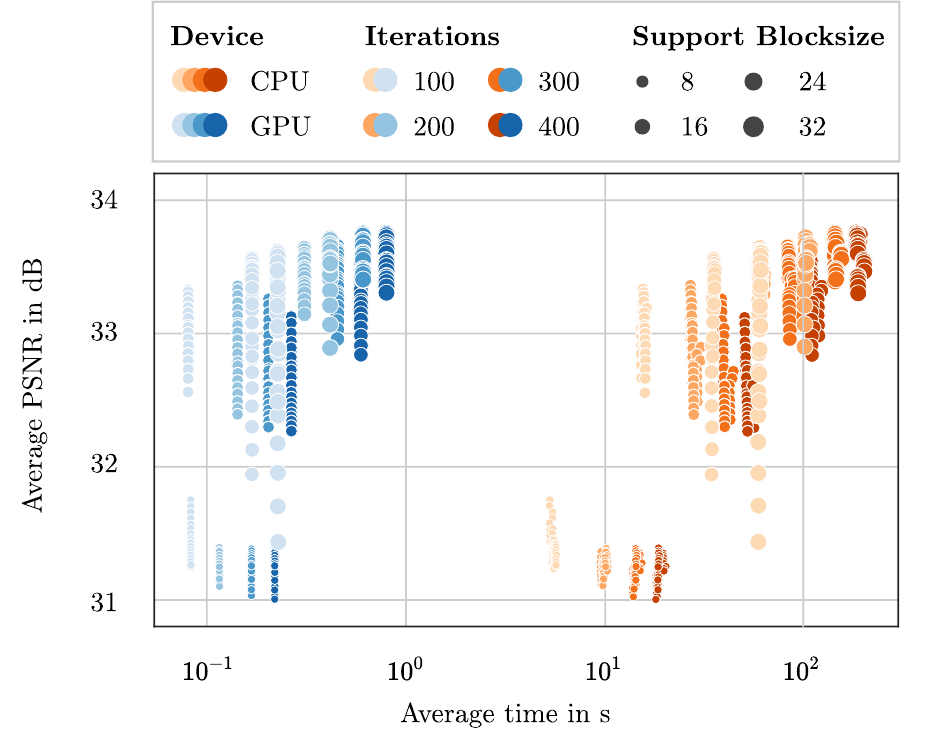}
\caption{Extensive parameter exploration regarding quality and execution time for the CPU and the GPU implementations of the \ac{FSR}. Each data point represents a single parameterization averaged over all test images. In total, 640 parameterizations were tested. For details, see text.}
\label{simulations:fig:parameter_sweep}
\end{figure}

As the proposed parallelization of \ac{FSR} follows the same mathematical procedure as the conventional \ac{FSR} on the CPU, both implementations produce identical results for a given parameterization. However, due to the massively parallel execution of \ac{FSR} on the GPU, a considerable gain of speed with respect to the single-threaded CPU implementation can be observed as visible in Fig. \ref{simulations:fig:time_gpu_cpu}. Thereby, results for a shared memory GPU implementation are included, where the argmax calculation is performed using a shared memory approach as described in \cite{Dai2014} instead of the novel register-based approach. Other than that, the shared memory GPU implementation is identical to the introduced parallelization of \ac{FSR}. It is clearly visible, that the register-based argmax calculation yields notable speed improvements over the shared memory approach. With the register-based argmax calculation, speed improvements of more \mbox{than $100\times$} are possible with respect to the single-threaded CPU implementation. While the CPU performance could be greatly improved by higher clock frequencies and performing the reconstruction of different support blocks in parallel on multiple CPU cores, note, that even for a twice as fast clock frequency, more \mbox{than 50 CPU} cores are required to compete with the performance of the highly parallelized \ac{FSR} on a single GPU.

Fig. \ref{simulations:fig:parameter_sweep} verifies that the significant gain of speed can be generalized to a wide range of parameters. Each data point represents the PSNR and execution time for a given parameterization averaged over all test images. The number of \mbox{iterations $I$} has been varied in the range $[100; 400]$ using a step size of $100$, the support blocksize $S$ has been varied in the range $[8; 32]$ using a step size of $8$, the spatial decay factor $\hat{\rho}$ has been varied in the range $[0.68; 0.82]$ using a step size of $0.02$, and the orthogonality deficiency compensation \mbox{factor $\gamma$} has been varied in the range $[0.2; 0.6]$ using a step size \mbox{of $0.1$}. Therefore, 640 parameterizations are tested for the GPU and the CPU implementation, each. A notable gain of speed can be observed for all parameterizations by employing the highly parallelized GPU implementation compared to its CPU counterpart. For all tested parameterizations, one can expect the GPU implementation to be roughly $100\times$ faster than the single-threaded CPU implementation, validating our findings from above.

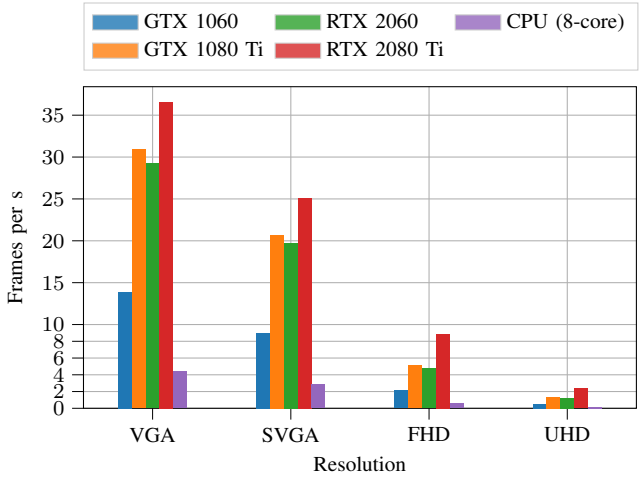
\begin{figure}[t]
	\setlength\figurewidth{\columnwidth}
	\setlength\figureheight{.66\columnwidth}
	\centering
	{\footnotesize\input{tikz/multiresolution}}
	\caption{Average framerate of the novel parallelized FSR for various resolutions on different GPUs compared to a multithreaded CPU implementation on an Intel Core\textsuperscript{\texttrademark} i9-9900 @ 3.1 GHz.}
	\label{simulations:fig:multiresolution}
\end{figure}

With many parameterizations of the parallelized \ac{FSR} finishing considerably below one second while still producing decent results in terms of quality, a look at the achievable framerates measured in \ac{fps} is of interest. \mbox{Fig. \ref{simulations:fig:multiresolution} compares} the average framerate of the parallelized \ac{FSR} for various common video resolutions on different GPUs to a multithreaded implementation of \ac{FSR} on the CPU.
Thereby, an \mbox{Intel Core\textsuperscript{\texttrademark} i9-9900 @ 3.1 GHz} is employed for the CPU computations, where the 8 cores of the CPU perform the reconstruction of different support blocks simultaneously.
The investigated video resolutions are \mbox{$640 \times 480$ (VGA)}, \mbox{$800 \times 600$ (SVGA)}, \mbox{$1920 \times 1080$ (FHD)} and \mbox{$3840 \times 2160$ (UHD)}. The support blocksize is set \mbox{to $S = 16$}. Considering VGA, more than $30$ fps can be achieved which makes the proposed highly parallelized \ac{FSR} capable of real-time applications. Furthermore, it is visible that the algorithm scales reliably with the computing power of different GPUs. For the larger FHD resolution, more \mbox{than $1$ fps} is achieved on all applied GPUs, while the powerful \mbox{RTX 2080 Ti} reaches almost \mbox{$10$ fps}. Even for the state-of-the-art UHD resolution, most of the GPUs among the test field are still able to steadily reconstruct the original image data in less than one second. The multithreaded CPU implementation reaches more than $1$ fps only for the low resolutions VGA and SVGA, and, with a loss of about $70\%$ in terms of framerate, is well behind even the \mbox{slow GTX 1060} for all investigated resolutions.

%% file: tikz/time_gpu_cpu.tex
\begin{tikzpicture}

\definecolor{color0}{rgb}{1,0.498039215686275,0.0549019607843137}
\definecolor{color1}{rgb}{0.172549019607843,0.627450980392157,0.172549019607843}
\definecolor{color2}{rgb}{0.12156862745098,0.466666666666667,0.705882352941177}

\begin{axis}[
height=\figureheight,
legend cell align={left},
legend style={fill opacity=0.8, draw opacity=1, text opacity=1, at={(0.03,0.97)}, anchor=north west, draw=white!80!black},
log basis y={10},
tick align=outside,
tick pos=left,
width=\figurewidth,
x grid style={white!69.0196078431373!black},
xlabel={Support blocksize \(\displaystyle S\)},
xmajorgrids,
xmin=2.6, xmax=33.4,
xtick={4,8,16,24,32},
xtick style={color=black},
y grid style={white!69.0196078431373!black},
ylabel={Time in s},
ymajorgrids,
ymin=0.0419238898096356, ymax=87.7472491539009,
ymode=log,
ytick style={color=black}
]
\addplot [ultra thick, color0, mark=*, mark size=2, mark options={solid}]
table {%
4 2.47513045787811
8 5.83487643003464
16 16.375654668808
24 36.2915032172203
32 61.9853890371323
};
\addlegendentry{CPU}
\addplot [ultra thick, color1, mark=*, mark size=2, mark options={solid}]
table {%
4 0.0967297840118408
8 0.12172425031662
16 0.196670703887939
24 0.324537065029144
32 0.538598194122314
};
\addlegendentry{GPU (shared)}
\addplot [ultra thick, color2, mark=*, mark size=2, mark options={solid}]
table {%
4 0.0593479537963867
8 0.0631702542304993
16 0.0804236245155334
24 0.18246833562851
32 0.227635178565979
};
\addlegendentry{GPU (ours)}
\end{axis}

\end{tikzpicture}

%% file: tikz/multiresolution.tex
\begin{tikzpicture}

\definecolor{color0}{rgb}{0.12156862745098,0.466666666666667,0.705882352941177}
\definecolor{color1}{rgb}{1,0.498039215686275,0.0549019607843137}
\definecolor{color2}{rgb}{0.172549019607843,0.627450980392157,0.172549019607843}
\definecolor{color3}{rgb}{0.83921568627451,0.152941176470588,0.156862745098039}
\definecolor{color4}{rgb}{0.580392156862745,0.403921568627451,0.741176470588235}

\begin{axis}[
height=\figureheight,
legend columns=3,
legend cell align={left},
legend style={fill opacity=0.8, draw opacity=1, text opacity=1, draw=white!80!black, at={(0, 1.05)}, anchor=south west},
tick align=outside,
tick pos=left,
width=\figurewidth,
x grid style={white!69.0196078431373!black},
xlabel={Resolution},
xmajorgrids,
xmin=-0.5, xmax=3.5,
xtick style={color=black},
xtick={0,1,2,3},
xticklabels={VGA,SVGA,FHD,UHD},
y grid style={white!69.0196078431373!black},
ylabel={Frames per s},
ymajorgrids,
ymin=0, ymax=38.4001750109682,
ytick={0,2,4,6,8,10,15,20,25,30,35},
ytick style={color=black}
]
\addlegendimage{area legend,draw=none,fill=color0};
\addlegendentry{\phantom{}GTX 1060\phantom{.}}
\addlegendimage{area legend,draw=none,fill=color2};
\addlegendentry{\phantom{}RTX 2060\phantom{.}}
\addlegendimage{area legend,draw=none,fill=color4};
\addlegendentry{\phantom{}CPU (8-core)}
\addlegendimage{area legend,draw=none,fill=color1};
\addlegendentry{\phantom{}GTX 1080 Ti\phantom{.}}
\addlegendimage{area legend,draw=none,fill=color3};
\addlegendentry{\phantom{}RTX 2080 Ti\phantom{.}}

\draw[draw=none,fill=color0] (axis cs:-0.25,0) rectangle (axis cs:-0.15,13.8522465855471);

\draw[draw=none,fill=color0] (axis cs:0.75,0) rectangle (axis cs:0.85,8.99940176210918);
\draw[draw=none,fill=color0] (axis cs:1.75,0) rectangle (axis cs:1.85,2.13324077133165);
\draw[draw=none,fill=color0] (axis cs:2.75,0) rectangle (axis cs:2.85,0.541409823725059);
\draw[draw=none,fill=color1] (axis cs:-0.15,0) rectangle (axis cs:-0.05,30.8955359395083);

\draw[draw=none,fill=color1] (axis cs:0.85,0) rectangle (axis cs:0.95,20.6951966211548);
\draw[draw=none,fill=color1] (axis cs:1.85,0) rectangle (axis cs:1.95,5.12690055882142);
\draw[draw=none,fill=color1] (axis cs:2.85,0) rectangle (axis cs:2.95,1.38327128913797);
\draw[draw=none,fill=color2] (axis cs:-0.05,0) rectangle (axis cs:0.05,29.2407481157131);

\draw[draw=none,fill=color2] (axis cs:0.95,0) rectangle (axis cs:1.05,19.7101144873555);
\draw[draw=none,fill=color2] (axis cs:1.95,0) rectangle (axis cs:2.05,4.75572894638932);
\draw[draw=none,fill=color2] (axis cs:2.95,0) rectangle (axis cs:3.05,1.19513968629171);
\draw[draw=none,fill=color3] (axis cs:0.05,0) rectangle (axis cs:0.15,36.5715952485412);

\draw[draw=none,fill=color3] (axis cs:1.05,0) rectangle (axis cs:1.15,25.1333718115556);
\draw[draw=none,fill=color3] (axis cs:2.05,0) rectangle (axis cs:2.15,8.85813488631055);
\draw[draw=none,fill=color3] (axis cs:3.05,0) rectangle (axis cs:3.15,2.36959507468712);
\draw[draw=none,fill=color4] (axis cs:0.15,0) rectangle (axis cs:0.25,4.39876597427769);

\draw[draw=none,fill=color4] (axis cs:1.15,0) rectangle (axis cs:1.25,2.84380966897864);
\draw[draw=none,fill=color4] (axis cs:2.15,0) rectangle (axis cs:2.25,0.589536201442546);
\draw[draw=none,fill=color4] (axis cs:3.15,0) rectangle (axis cs:3.25,0.146694319217363);
\end{axis}

\end{tikzpicture}

%% file: sections/conclusion.tex
\section{Conclusion}
\label{sec:conclusion}

In this paper, a highly parallellized implementation of \ac{FSR} carefully designed for the execution on GPUs has been introduced. In combination with the novel highly effective, high-speed argmax calculation based on direct register access between neighbouring threads, the execution time of the reconstruction procedure can be considerably reduced. The register-based argmax calculation proved to be a major accelerator of the overall reconstruction procedure leading to notable speed improvements over the shared memory based approach.
Depending on the applied hardware, speed-ups of more than $100\times$ are possible compared to previous approaches without losing image quality. These speed-ups are reliably obtained for a wide range of parameterizations.
All in all, the proposed highly parallelized approach is able to reconstruct multiple frames per second on a wide range of hardware and is capable of real-time applications with more than $30$ frames per second.